\documentclass[10pt, twocolumn, a4page]{ileapr}
\usepackage{graphicx}
\usepackage{times}    
\usepackage{amsmath}  
 \newcommand{\mb}[1]{\mathbf{#1}}
 
 \newcommand{\mr}[1]{\mathrm{#1}}

 \newcommand{\D}{\Delta}
\input{ileapr.mac}
\title{
 \vskip -15mm
         {\large Annual Progress Report 2003, Institute of Laser Engineering, Osaka University (2004), pp.147-150}\\
 \vskip -5mm
\rule{170mm}{0.2mm}
 \vskip  2mm
{\bf A New Dynamical Domain Decomposition Method for Parallel Molecular Dynamics Simulation on Grid}}
\author{ Vasilii Zhakhovskii, Katsunobu Nishihara, Yuko Fukuda, and Shinji Shimojo}
 \date{ {\small
                                                                 } }

\begin{document}
\maketitle
\thispagestyle{empty}

   \section*{INTRODUCTION}
   \vspace{-2mm}

Classical molecular dynamics (MD) approach is widely applied for simulation of complex many-particle systems in many areas of physics, chemistry and biochemistry. The particles such as atoms and molecules interact with each other through a given force function. By step-by-step integration of Newton's equations of motion, trajectories of all particles are obtained.  Useful information can be extracted from the particle trajectories by using a suitable averaging procedure.

Typically the interactions among the particles can be described by short-range forces, hence MD simulation has local spatial character. Therefore the particle motion for a short period can be determined only from particles located in neighborhood of a given particle. This fact was used to develop successful parallel MD algorithms based on a static spatial domain decomposition (SDD) of simulation area \cite{Plimpton95}. There are a few approaches to improve the load balancing of SDD \cite{Nyland97},\cite{Deng00}, which works well for simulations with uniform density distribution and without significant flows of particles.

A large scale MD simulation requires a large number of high performance computers. Recently the Grid computing has been developed and allows us to use many computers connected through a net. However performance of the computers and the net may not be the same and their performance may not be known before the use. In addition, other users may also run their programs on the same computer network without notice. If we use many computers located in different sites, some of the computers may be very busy for certain times because of other users. A dynamical domain decomposition method is therefore required to obtain a good adaptive load balancing for heterogeneous computing environments such as Grid. We have developed a new Lagrangian material particle -- dynamical domain decomposition method ($\mr{MPD}^3$) for large scale parallel MD simulation on a heterogeneous computing net, which results in large reduction of elapsed time.

Recently $\mr{MPD}^3$ method was applied for simulations of hydrodynamics problems like the Richtmyer-Meshkov instability \cite{IFSA2001} and laser destruction of solids \cite{Anisimov03}. The flows of matter in both cases result in great imbalance among processors in SDD. The new method is also applicable for simulations of any dynamical processes with strongly nonuniform density distribution accompanied with phase transition, shock waves and cracks.

In this report we present a new $\mr{MPD}^3$ algorithm and its performance tested for real physical problems in various computing environments, such as PC clusters connected within LAN and super computers connected with Super SINET, with the use of Globus toolkit 2.0.

 \begin{figure}
  \begin{center}
    \includegraphics [width=0.65\columnwidth] {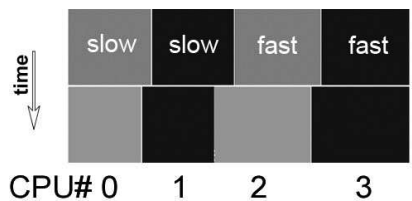}
    \caption{ \label{fig:4CPU} 1D dynamical domain decomposition for simulation of a crystal bar at rest
              with the use of 2 slow CPUs (0,1 Xeons) and 2 fast CPUs (2,3 AMD) connected with 100 Mbps LAN.}
 \end{center}
  \vspace{-5mm}
 \end{figure}

 \vspace{-1mm}
 \section*{ALGORITHM}
 \vspace{-2mm}

Let us consider many particles interacting each other with short-range forces in a MD simulation box that is divided into subdomains. MD simulation is carried out using many processors connected through nets, where a number of the processors is $N_p$. Each processor calculates MD dynamics of the particles belonging to a subdomain. We consider following computing environment: other user's programs and system programs are running on some of the $N_p$ processors and a number of other programs may change during the simulation. Our main goal is to develop successful algorithm to accomplish a good load balance among the processors to reduce elapsed time. It should be noted that the computational loading of each processor may also change in time due to the dynamical processes in the subdomains. Therefore the load balance algorithm has to be iteratively adapted for time-dependent computing environment and simulation dynamics.


 \begin{figure}[ht]
 \begin{center}
    \includegraphics [width=\columnwidth] {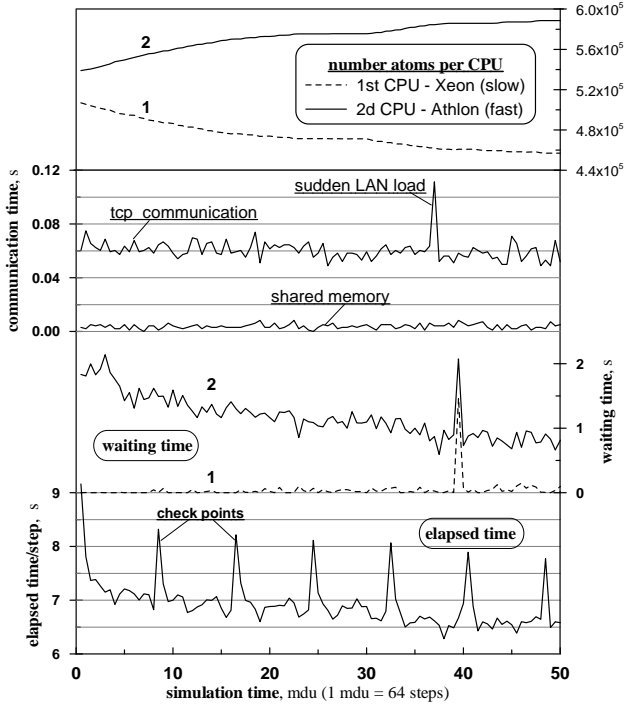}
    \caption{ \label{fig:4CPU-time} Adaptation of the $\mr{MPD}^3$ method to the imbalanced heterogeneous computing
               net managed by MPICH for simulation of crystal in Fig.\ref{fig:4CPU}. Number of atoms per CPU,
               communication time, waiting time and elapsed time per step are shown as functions of simulation steps
               from the top to the bottom. Check points denote massive hard disk operations. }
 \end{center}
  \vspace{-5mm}
 \end{figure}

We assume here that each simulation step in the algorithm can be divided into two parts -- MD simulation of particle motion and exchange of particle data among the processors. A program may receive both the CPU-dependent time and the elapsed time for each part. Here the CPU-dependent time is time duration spent only for our simulation in each processor. If other programs run on some of the processors, the elapsed time becomes longer than the CPU-dependent time in those processors.

The normalized MD working time of the $i\!$-th processor $P(i)$ can be defined as
\vspace{-2mm}
\begin{equation}
         P(i) = \frac{t_{MD}(i)}{t_{eMD}(i)},  \qquad  0 < P(i) \le 1     \label{eq:alg1}
\vspace{-1mm}
\end{equation}
where $t_{MD}$ is the CPU-dependent time spent for the MD simulation without communication with other processors, and $t_{eMD}$ is the elapsed time for MD simulation. The $P(i)$ is useful to estimate how other programs are loaded on the  $i\!$-th processor. If there is no other program, it becomes one.
Let us define the weighting factor $W(i)$ of the $i\!$-th CPU as
\vspace{-2mm}
\begin{equation}
         W(i) = \frac{t_w(i)}{P(i)t_e(i)},  \qquad  0 < W(i) \le 1     \label{eq:alg2}
 \vspace{-1mm}
\end{equation}
Here $t_w$ is the CPU-dependent working time, and $t_e$ is the elapsed time, where both include time duration for the MD simulation and the communication. It is reasonable to suppose that the $i\!$-th CPU is in a good balance with the $j\!$-th CPU, if their weighting factors almost equal each other: $W(i) \cong W(j)$ .

A simulation domain can be divided into $N_p$ simple subdomains like rectangular boxes before the MD simulation, where $N_p$ equals a number of processors (CPUs). Each CPU calculates each sub-domain. $N(i)$ is a number of particles in the $i\!$-th subdomain. We define the position of the center $\mb{R}(i)$ of the $i\!$-th subdomain as following:
\vspace{-2mm}
\begin{equation}
         \mb{R}(i) = \frac{1}{N(i)} \sum_{k=1,N(i)} \mb{r}(k),     \label{eq:alg3}
\vspace{-1mm}
\end{equation}
where $k$ denotes a particle number in the $i\!$-th subdomain, $\mb{r}(k)$ is a position of the $k\!$-th particle. For identical particles, $\mb{R}(i)$ is a center of mass of the $i\!$-th subdomain.
For each pair of the $i,j\!$-subdomains, we define a boundary plane between them at the midpoint
$\mb{R}_{1/2}(i,j)= (\mb{R}(i)+\mb{R}(j))/2$ and perpendicular to the connecting vector
$\mb{R}(i,j)= \mb{R}(i)-\mb{R}(j)$. After that all of the particles near the boundary are associated with one of the $i,j\!$-subdomains. By repeating this procedure for all pairs, the simulation domain will be finally divided into steady $N_p$ Voronoi polygons. The map of Voronoi polygons is known as the Dirichlet tessellation and used for grid generation in computational fluid dynamics.

For a system with a uniform number density, the area of each Voronoi polygon is the same and thus the number of particles becomes the same in each domain. Therefore the Voronoi decomposition gives a good balance only for a homogeneous computing net and uniform density distribution. Nevertheless the Voronoi decomposition is a good point to start MD simulation. If the particle exchange and diffusion between the polygons are forbidden, the dynamical behavior of these polygons looks as motion of Lagrangian particles from the hydrodynamics point of view. This is the reason why we name the moving Voronoi polygon as a material particle (MP).

To obtain a good load balance in the heterogeneous computing environments the MP centers should be adjust
time-dependently so that for example a busy computer calculates less number of particles compared with others. We propose the following simple and efficient iterative algorithm to calculate a displacement of the MP center.
The displacement of the $i\!$-th MP center at the current $n$ simulation step can be evaluated as
\vspace{-2mm}
\begin{equation}
      \D \mb{R}^{n}(i) = \frac{a L(i)}{N_n(i)} \sum_{j=1,N_n(i)} (W(i)-W(j))
                           \frac{\mb{R}(i,j)}{|\mb{R}(i,j)|},      \label{eq:alg4}
\vspace{-1mm}
\end{equation}
where $N_n(i)$ is a number of neighbor MPs surrounding the $i\!$-th MP, $L(i)$ is a linear size of the $i\!$-th MP,
and $a \in [0,1]$ is an adjustable parameter of the method.
At the next MD step $(n+1)$ a new position of the $i\!$-th MP $\mb{R}^{n+1}(i)$ is given by
\vspace{-2mm}
\begin{equation}
      \mb{R}^{n+1}(i) = \mb{R}^{n}(i) + \D \mb{R}^{n}(i)    \label{eq:alg5}
\vspace{-1mm}
\end{equation}
It is clear from Eqs.(\ref{eq:alg4}) and (\ref{eq:alg5}) that a good load balance may not be reached within a few steps. In addition, the well-balanced Voronoi decomposition may not exist for small numbers of CPUs $N_p$.
Nevertheless, as observed in our simulations with not so many processors of $N_p =8,12,14,16$, the good-balanced decompositions are achieved within 10\% of imbalance $|W(i)-W(j)|$ between the fastest CPU and the slowest one.

The simplified $\mr{MPD}^3$ {\bf algorithm} can be designed as follows:
\begin{itemize}
    \item[0)] {\bf Initialization.} Simple initial domain decomposition.
    \item[1)] Exchange particles between neighbor MPs according to the Voronoi method.
    \item[2)] Evaluate the new MP positions and iterate 1)-step so long as every MPs reach steady shapes.
    \item[3)] {\bf Start simulation.} Exchange the MP positions, timing data, particles and etc. among the neighbor MPs. Evaluate a new desired position of the MP center by using Eqs.(\ref{eq:alg4}) and (\ref{eq:alg5}).
    \item[4)] Advance MD integration step and measure all timing data:
              $t_{MD}(i),t_{eMD}(i),t_w(i),t_{e}(i)$.
    \item[5)] {\bf Repeat } 3)- and 4)-steps.
\end{itemize}

It should be pointed out that the new $\mr{MPD}^3$ method shares features with the usual spatial decomposition method as well as the particle decomposition algorithm \cite{Plimpton95}. In other words the simulated particles are distributed among CPUs in according to their positions in the dynamical clustered medium of subdomains/MPs which depend on particle motion. A particle keeps its number and membership of MP only for a relatively short period. To prevent the increase of cache memory missing we renumber particles inside MP in some geometrical order, for instance by using a rectangular mesh. It has been observed in our MD simulations that the renumbered neighbor particles lie short distances from each other in computer memory on average. The renumbering significantly improves cache hitting and CPU performance.

 \vspace{-2mm}
 \section*{TESTING}
 \vspace{-1mm}

The $\mr{MPD}^3$ method was implemented in the full vectorized Fortran program with calls of standard MPI 1.1 subroutines. The $\mr{MPD}^3$ program was tested in various computing environments and physical problems.
We measured performance of the $\mr{MPD}^3$ method in details for two testing problems, namely 1D and 2D decompositions, which demonstrated applicability of the new method to computer clusters and Grid computing.

The first simple test is 1D decomposition for MD simulation of a steady crystal bar at rest with the use of PC clusters consisting of different cpu performances connected within LAN. There are two dual-processor computers, 2-Xeon 2.2 GHz and 2-AMD 1.5 GHz, connected by 100 Mbps LAN, namely 4 CPUs in total. Communication is managed by the well known free MPICH software (see http://www-unix.mcs.anl.gov/mpi/mpich/ ). This model task is expected to check adaptability of the $\mr{MPD}^3$ to the heterogeneous computing net. Figure \ref{fig:4CPU} shows the initial decomposition on the top and the well-balanced final decomposition on the bottom. The corresponding timing data are presented in Fig.\ref{fig:4CPU-time}. On the top, it can be seen how the MPs belonging to either fast or slow CPUs exchange their particles among them to achieve a good load balance. The second graph from the top shows communication time between CPUs. Share memory communication is the superior, and the line with the bandwidth of 100 Mbps also demonstrates very good performance of which duration is less than 1\% of the ealpsed time shown in the bottom of Fig.\ref{fig:4CPU-time}. It is interesting to note that the method is inert to a short random load during few tens of simulation steps as shown around 40 mdu.
Due to the adaptation process in the $\mr{MPD}^3$ method the waiting time is reduced to half of its initial value at the simulation step of 50 in MD unit as shown in the 3rd graph from the top. The $\mr{MPD}^3$ method results in the reduction of the elapsed time from 9.1 sec/step to 6.5 sec/step.

The second test deals with MD simulations of high-speed collision of two solid cylinders with different radii
with the use of the $\mr{MPD}^3$ method. It is clear that the static domain decomposition may have a hard load imbalance because of large flow velocity of the particles. We used two super computers, NEC SX-5 of Cybermedia Center at Osaka University and SX-7 of Information Synergy Center at Tohoku University. They are connected with Super-SINET with the use of Globus toolkit 2.0.

\begin{figure}[h]
 \begin{center}
    \includegraphics [width=0.9\columnwidth] {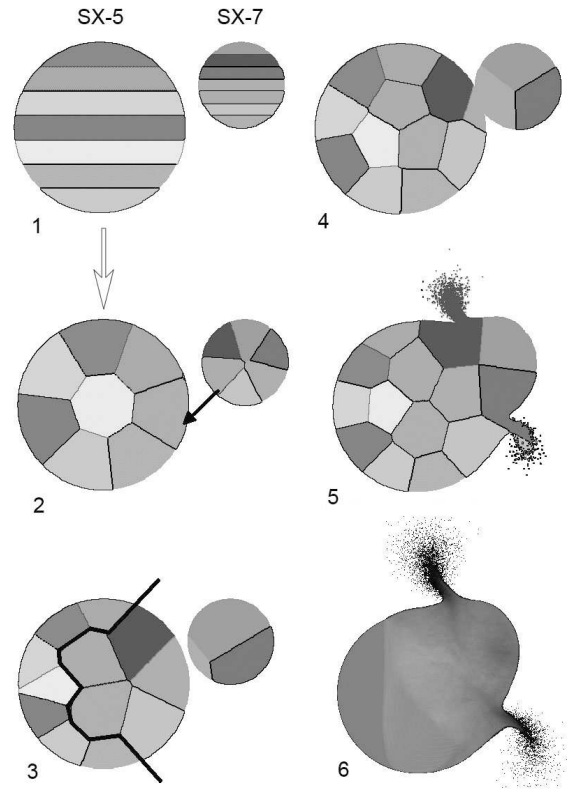}
    \caption{ \label{fig:sx5-sx7} Snapshots of the collision of two cylinders.
      {\bf 1.} Simple initial rectangular decomposition among SX-5 machine (left cylinder, 7 MPs) and SX-7
               machine (right cylinder, 7 MPs).
      {\bf 2.} Start point for MD simulation. Steady Voronoi decomposition was achieved after 14 iterations. One MP
               moves from right cylinder to the left as indicated by arrow.
      {\bf 3.} Black line shows the boundary between overloaded SX-5 and low loaded SX-7 machines
               at the end of the (1)st period shown in Fig.\ref{fig:sx5-sx7_time}.
      {\bf 4.} Both computers work in exclusive usage mode which corresponds to the (2)nd period in
               Fig.\ref{fig:sx5-sx7_time}. MP subdomains are almost equal in size.
      {\bf 5.} Collision results in adaptive deformation of decomposition at the (4)th period in
               Fig.\ref{fig:sx5-sx7_time}.
      {\bf 6.} Mass density map of two colliding bodies at the end of simulation.
               Gray scale indicates density, i.e., lighter gray denotes more dense material.  }
 \end{center}
  \vspace{-5mm}
 \end{figure}

Figure \ref{fig:sx5-sx7} shows a set of snapshots from the beginning to the end of simulation. Initially the simplest rectangular decomposition is chosen for dividing simulation domain among 7 SX-5 CPUs and 7 SX-7 CPUs as shown in Fig.\ref{fig:sx5-sx7}.1. After 14 iterations, the steady Voronoi decomposition is established. At the moment the left cylinder consists of 8 MPs, of which 7 MPs belong to SX-5 and other 1 MP to SX-7, and the right cylinder consists of 6 MPs belonging to SX-7, see Fig.\ref{fig:sx5-sx7}.2. As it has been already pointed out this is the starting point of the MD simulation. After that the performance of each CPU is measured.


  \begin{figure*}[ht]
 \begin{center}
    \includegraphics [width=0.72\textwidth] {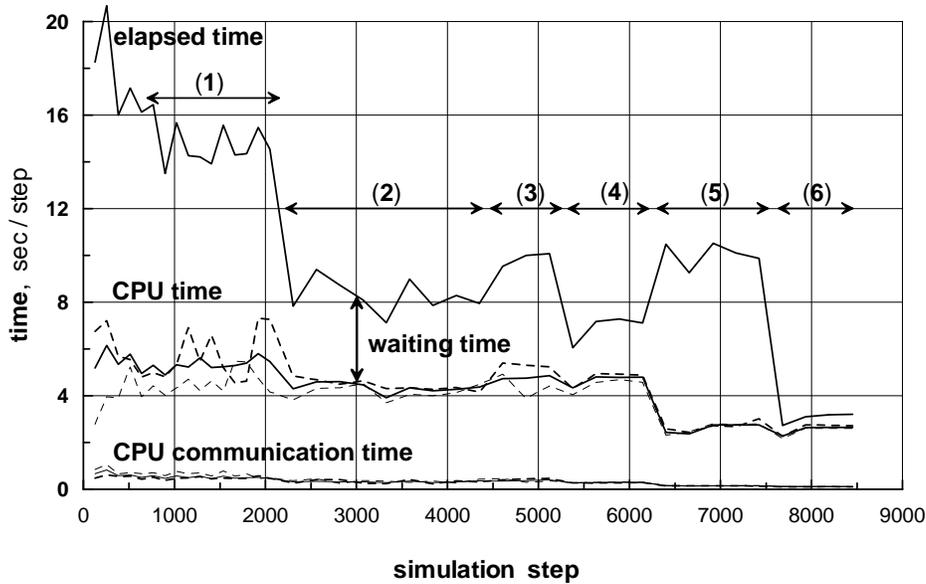}
    \caption{ \label{fig:sx5-sx7_time} Performance of the $\mr{MPD}^3$ method for simulation of two bodies collision (see Fig.\ref{fig:sx5-sx7}) in different computer environments including the Grid environment on NEC SX-5 (Osaka University) and NEC SX-7 (Tohoku University) machines connected with Super-SINET (see below for details). Waiting time indicates time duration for all MPI\_WAIT subroutines per one simulation step. CPU communication time denotes a CPU-dependent time of communication subprograms including all MPI\_ISEND, MPI\_IRECV and preparation                     operations like seeking, sorting and so on.
  {\bf (1).} Adaptation of the $\mr{MPD}^3$ method to SX-5 overloaded by other users and almost free SX-7 computer,
             see snapshot Fig.\ref{fig:sx5-sx7}.3.
  {\bf (2).} Both computers work in exclusive usage mode corresponding to Fig.\ref{fig:sx5-sx7}.4. Long waiting time
             is due to the low bandwidth of the network line between two computers.
  {\bf (3).} Simulation in the Grid on two nodes of overloaded SX-5 connected with Giga-bit Ether.
  {\bf (4).} The same as (3), but both nodes work in exclusive usage mode, see snapshot in
             Fig.\ref{fig:sx5-sx7}.5. Still the low bandwidth line between the nodes results in the long waiting time.
  {\bf (5).} Simulation within a single node of SX-5 overloaded by other users, but communication
             through share memory only.
  {\bf (6).} The same as (5), but SX-5 works in exclusive usage mode.
             The waiting time is reduced to $\sim 20\%$ of CPU-dependent time.         }
 \end{center}
  \vspace{-5mm}
 \end{figure*}

The $\mr{MPD}^3$ algorithm shall attempt a load balance between  CPUs. In Fig.\ref{fig:sx5-sx7}.3, the thick black line denotes the boundary between SX-5 and SX-7 at the end of the 1st period corresponding to (1) in Fig.\ref{fig:sx5-sx7_time}. At this time SX-5 at Osaka University is overloaded by other users while SX-7 at Toholu University is less loaded. As indicated in the figure, it results in reduction of the areas corresponding to the MPs belonging to SX-5.  The two neighbor CPUs can be in balance by giving the particles from busy CPU to free one.

The snapshot in Fig.\ref{fig:sx5-sx7}.4 shows more or less uniform distribution in MP sizes. It reflects that the
simulation is carried out in exclusive usage mode on both computers as shown in (2) Fig.\ref{fig:sx5-sx7_time}.
The adaptation from Fig.\ref{fig:sx5-sx7}.3 to Fig.\ref{fig:sx5-sx7}.4 states takes about 500 simulation steps.
The next snapshot in Fig.\ref{fig:sx5-sx7}.5 shows the adaptive deformation of the domains due to two body collision. It corresponds to the end of the (4)th period in Fig.\ref{fig:sx5-sx7_time} when the simulation has been performed in local Grid environment on the two different nodes of the NEC SX-5 at Osaka University connected with Giga bit ether.
The low bandwidth line between the nodes results in the long waiting time still. Fig.\ref{fig:sx5-sx7}.6 shows a mass density map of two bodies at the end of the simulation.

At the beginning of the (5)th period in Fig.\ref{fig:sx5-sx7_time} the simulation program was recompiled for standard MPI environment on single node of SX-5 machine. It results in a pronounced reduction of CPU-dependent time. During the (6)th period in Fig.\ref{fig:sx5-sx7_time}, SX-5 works in exclusive usage mode. The waiting time becomes near $\sim 20\%$ of the CPU time/step and can not be decreased further. Most probably the number of CPUs is too small to optimize the MP sizes in the frame of Voronoi decomposition.

  \vspace{-1mm}
  \section*{ CONCLUSION }
  \vspace{-1mm}

  We have demonstrated that the $\mr{MPD}^3$ method is a highly adaptive dynamic domain decomposition algorithm for MD simulation on both PC clusters and Grid computing environments, even if other programs are running on the same environment. It has been shown that the well-balanced decomposition results from dynamical Voronoi polygon tessellation, where each polygon is considered as a material or Lagrangian particle and its center is displaced to reach the minimum elapsed time with a good load balance. Our approach can be extended to other particle methods like Monte Carlo, particle-in-cell, and smooth-particle-hydrodynamics.

The $\mr{MPD}^3$ method works perfectly for 1D decomposition, but for 2D case the load balance may depend on a geometrical configuration of simulation problems and a number of CPUs in use. We suppose that the $\mr{MPD}^3$ is especially optimal for large scale simulation with a large number of computers.

Although we do not check parallel efficiency (scalability) of the $\mr{MPD}^3$ method, it must be nearly the same  ($\!\sim 90\%$) as the SDD method on the uniform medium \cite{Plimpton95}.

We are grateful to the Cybermedia Center at Osaka University and Information Synergy Center at Tohoku University for the organization of computer experiments.



\vspace{-1mm}
\begin{thebibliography}{9}
\vspace{-0mm}

\bibitem{Plimpton95} S.Plimpton, {\it J. Comput. Phys.} {\bf 117},1,(1995)
\vspace{-2mm}
\bibitem{Nyland97}   L.Nyland et al, {\it J. Parallel and Distributed Computing} {\bf 47},125,(1997)
\vspace{-2mm}
\bibitem{Deng00}     Y.Deng, R.F.Peierls, and C.Rivera, {\it J. Comput. Phys.} {\bf 161},1,(2000)
\vspace{-2mm}
\bibitem{IFSA2001}  V.~Zhakhovskii, K.~Nishihara, and M.~Abe in {\it Inertial Fusion Science and Applications,
                    IFSA 2001}, (Elsevier, Paris, 2002), pp.106-110
\vspace{-2mm}
\bibitem{Anisimov03} S.I.Anisimov, V.V.Zhakhovskii, N.A.Inogamov, K.Nishihara, A.M.Oparin, and Yu.V.Petrov
                     JETP Lett. {\bf 77}, 606 (2003), and see the previous report as well.

\end{thebibliography}
\end{document}